\documentclass[preprint,12pt]{elsarticle}

\usepackage{amssymb}
\usepackage{mathrsfs}

\usepackage{graphicx}
\usepackage{dcolumn}
\usepackage{bm}

\journal{Physics Letter B}

\begin{document}

\begin{frontmatter}



\title{Is a co-rotating Dark Disk a threat to Dark Matter Directional Detection ?}


\author{J. Billard, Q. Riffard, F. Mayet, D. Santos}

\address{Laboratoire de Physique Subatomique et de Cosmologie, Universit\'e Joseph Fourier Grenoble 1,
  CNRS/IN2P3, Institut Polytechnique de Grenoble, Grenoble, France}

\begin{abstract}
Recent N-Body simulations are in favor of the presence of a co-rotating Dark Disk that might contribute significantly 
(10\%-50\%) to the local Dark Matter density. Such substructure could have dramatic effect on directional detection. 
Indeed, in the case of a null lag velocity, one expects an isotropic WIMP velocity distribution
arising from the Dark Disk contribution, which might weaken the strong angular signature expected in directional detection. 
For a wide range of Dark Disk parameters, we evaluate in this Letter the effect of such dark component on the discovery potential 
of upcoming  directional detectors. As a conclusion of our study, using only the angular distribution of nuclear recoils, we show that Dark
Disk models as suggested by recent N-Body simulations will not affect significantly the Dark Matter reach 
of directional detection, even in extreme configurations. 
\end{abstract}


\end{frontmatter}


%
%
 
\newpage

\section{Introduction}
Within the standard Dark Matter halo paradigm, the local Dark Matter distribution is assumed to be smoothly  spatially distributed and 
to be well-described by a Maxwellian velocity distribution. However, the hierarchical structure formation model indicates that 
the Galactic Dark Matter halo results from successive small halo accretions, thus directly linking its structure to its merging
history. 
The presence of substructures in the Milky Way halo is inferred from recent results of N-body
simulation \cite{nbody.vl2,nezri,Kuhlen:2012fz,lisanti1,vogel,moore,klypin,kravtos}. Such substructures may be classified as follows : 
Dark Matter tidal streams (spatially localized), debris flows (spatially homogenized but with velocity substructures) 
and a Dark Disk. The latter has received much interest since 
late sub-halo merging is expected to lead to the formation of a co-rotating Dark Disk \cite{nezri,Read:2008fh,Read:2009,Purcell:2009yp} that may affect the expected WIMP signal both in
direct and directional detection.    
While the influence of the dark disk on Dark Matter signals
has been exhaustively investigated for direct \cite{nezri,Ling:2009cn,Bruch:2009rp,Green:2010gw} and indirect \cite{Bruch:2008rx} detection, it is still unclear
how it may affect directional detection. Following a previous work from A.~M.~Green \cite{Green:2010gw}, we aim at evaluating the 
influence of the presence of a co-rotating Dark Disk on the discovery potential of a forthcoming directional detector. 
In particular, we are interested in determining the values of the Dark Disk parameters for it to significantly affect the 
Dark Matter reach of directional detection. In order to be model independent from the background energy modelling, the study has been done by considering
 only the angular distribution of nuclear recoils $dR/d\Omega_r$.\\
Since the pioneering paper of D.~N.~Spergel~\cite{spergel}, the contribution of 
directional detection to the field of Dark Matter has been addressed through a wealth of 
studies~\cite{Kuhlen:2012fz,billard.exclusion,henderson,morgan1,morgan2,copi1,copi2,copi3,green1,green2,billard.disco,billard.profile,green.disco,billard.ident,Alves:2012ay,Lee:2012pf,albornoz,Bozorgnia:2011vc,Bozorgnia:2012,Creswick:2010dm,Lisanti:2009vy,Alenazi:2007sy,Gondolo:2002np}.
 Depending on the unknown WIMP-nucleon cross section, directional detection may be used to : 
exclude Dark Matter \cite{billard.exclusion,henderson}, reject the isotropy 
hypothesis \cite{morgan1,morgan2,copi1,copi2,copi3,green1,green2}, discover galactic Dark Matter with a high 
significance \cite{billard.disco,billard.profile,green.disco} or constrain WIMP and halo 
properties \cite{billard.ident,Alves:2012ay,Lee:2012pf}. In particular, for neutralino Dark Matter, a large fraction of MSSM configurations with a 
neutralino lighter than 200 $\rm GeV/c^2$ would lead  to a significance greater 
than 3$\sigma$ (90\% CL) in a 30 kg.year $CF_4$ directional detector \cite{albornoz}.\\
In the following, we focus on the effect of a co-rotating Dark Disk on the potential of 
forthcoming directional detectors to discover Dark Matter \cite{billard.profile}. 
The paper is organized as follows. Section  \ref{sec:dd} presents the current knowledge on the Dark Disk. In particular, we define  
its parameterization used throughout. The directional framework  is recalled in sec.~\ref{sec:directional}, with emphasize on the
directional statistic methods used to exploit forthcoming data. 
Then, for a wide range of Dark Disk parameters, we evaluate in \ref{sec:disco} the effect of such substructure  on the discovery potential 
of upcoming  directional detectors.

\section{A Dark Disk in the Milky Way}
\label{sec:dd}
Recent results from N-body simulation of Milky Way type galaxies have shown 
that merging satellite galaxies may get dragged into the plane of their host
galaxy \cite{nezri,Read:2008fh,Read:2009,Purcell:2009yp}. This leads to a Dark Matter overdensity roughly matching the
baryonic disk of the host galaxy and  usually in co-rotation with the latter \cite{Read:2008fh,Read:2009}.  This Dark Matter component is usually referred to as Dark Disk (DD). 
Hitherto, there is no observational evidence in favor of a Dark Disk in the Milky Way.\\ 
The Dark Disk is generally considered as a cold substructure for which the velocity distribution is described by an 
isotropic Maxwellian distribution \cite{Read:2008fh}. In such context, the astrophysical parameters relevant to the description of the Dark Disk are 
the density $\rho_{DD}$, its co-rotational velocity
$V_{DD}$ and its velocity dispersion given by $\sigma_{DD}$.\\
The range of interest of these parameters must be inferred from the results of N-body simulations and compared to astrophysical constraints. 
For instance, F.~S.~Ling {\it et al.} \cite{nezri} have extracted a Milky way 
type galaxy from the results of the RAMSES simulation \cite{nbody.Teyssier}. 
The velocity distribution of Dark Matter particles within a $7<R<9$ kpc and $|Z|<1$ kpc is shown to be well fitted
by a double Gaussian along the $\phi$ direction. The first one, corresponding to the Dark Matter halo component, 
is described by a null average speed and a velocity
 dispersion  $\sigma_{halo} \simeq 180$ km/s. The second component,
the Dark Disk, is described by a co-rotation velocity  $V_{DD} \simeq 150$ km/s and a velocity dispersion $\sigma_{DD} \simeq 85$ km/s. 
Note that this description is in good agreement with  \cite{Read:2008fh,Read:2009,Purcell:2009yp}. However, as there is 
no clear observational constraints on these  
parameters, we  allow for a wide range in order to
investigate the effect of a Dark Disk component on directional detection. Unless otherwise stated, we consider hereafter 
the following Dark Disk parameter ranges: 
\begin{eqnarray}
 0 <  & \rho_{DD}/\rho_{H}  & < 1  \nonumber \\ 
 0 \ \textrm{km.s$^{-1}$} <  & V_{DD}  & < 220 \ \textrm{km.s$^{-1}$}  \nonumber \\
 7 \ \textrm{km.s$^{-1}$} <  & \sigma_{DD}  & < 155 \ \textrm{km.s$^{-1}$} 
 \label{eq:ddparam}
\end{eqnarray}
Note that the dark disk properties depend on the merger history. For instance, and as outlined in \cite{Read:2008fh}, the dark disk density in the solar neigborhood could range between 20 and 100 per cent of
the halo one, for a given realization of Milky way like galaxies.
Any deviations from pure Gaussian and Maxwellian distributions, either for the halo or the Dark Disk component, may be treated  as in \cite{nezri} by using a 
generalized Maxwellian or a Tsallis distribution. However, as a simplifying assumption, although the real situation might
be more complicated, we will mostly consider an isotropic Maxwellian distribution to allow comparison with previous  works \cite{Green:2010gw} and discuss the case of an anisotropic Dark Disk velocity distribution at the end of section~\ref{sec:disco}. 

\section{Directional detection framework}
\label{sec:directional}
 \begin{figure*}[t]
\begin{center}
\includegraphics[scale=0.5,angle=0]{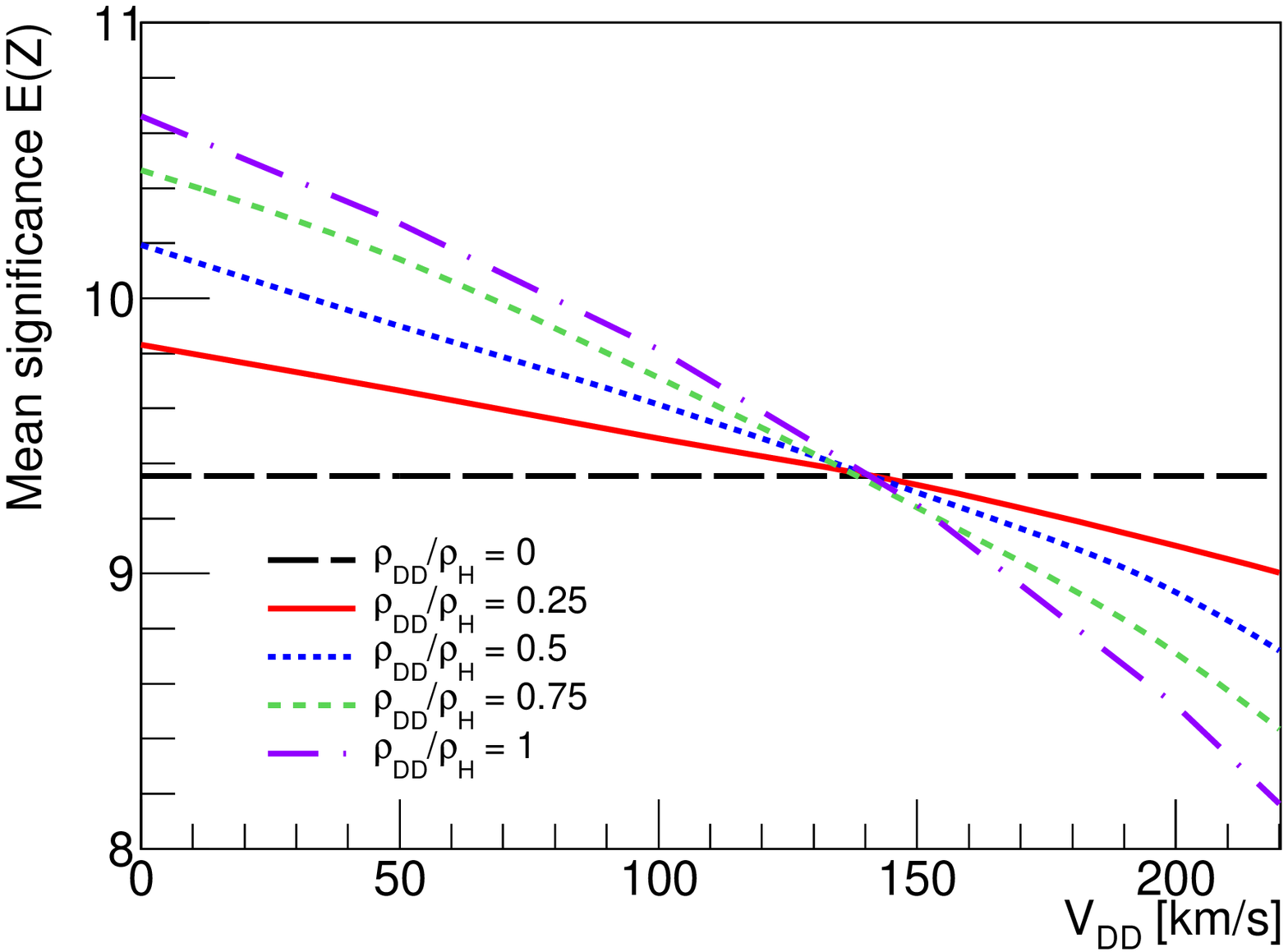}
\includegraphics[scale=0.5,angle=0]{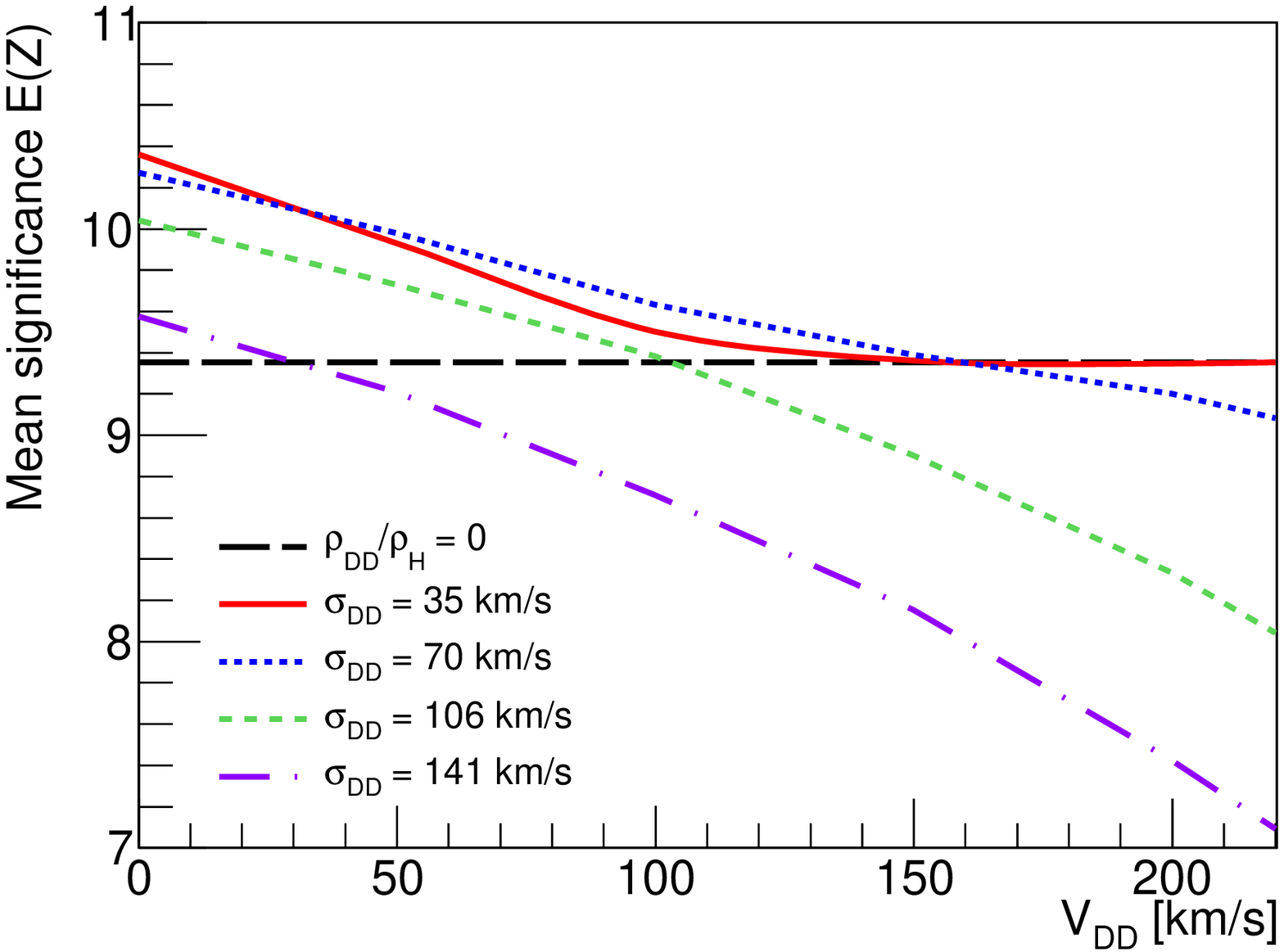}
\caption{Mean significance E(Z) as a function of $V_{DD}$, the rotation velocity of the Dark Disk at Solar radius.
Left : the result is presented for various values relative density 
$\rho_{DD}/\rho_{H}$ and a fixed velocity dispersion $\sigma_{DD} = 85$ km/s. Right : the result is presented for various values 
of the velocity dispersion  $\sigma^{DD}$ and a fixed value of the relative density $\rho_{DD}/\rho_{H} = 0.5$. 
These studies has been done for 50 GeV/c$^2$ WIMP mass and a  fixed value of 100  WIMP events (from the halo and the Dark Disk).} 
\label{fig:dispersion}
\end{center}
\end{figure*} 

There is a worldwide effort toward the development of a large TPC (Time Projection Chamber) devoted to directional detection 
\cite{white,dmtpc,mimac,newage,drift}. In the following, we
exemplify our Dark Disk study by considering a MIMAC-like detector corresponding to a low exposure (30 kg.year)  $\rm CF_4$ TPC allowing three dimensional track measurements \cite{mimac}.\\

The  two dimensional directional recoil  rate  $d^2R/dE_rd\Omega_r$ is given by  \cite{Gondolo:2002np} :
\begin{equation}
\frac{\mathrm{d}^2R}{\mathrm{d}E_r\mathrm{d}\Omega_r} = \frac{\rho_0\sigma_0}{4\pi m_{\chi}m^2_r}F^2(E_r)\hat{f}(v_{\textrm{min}},\hat{q}),
\label{directionalrate}
\end{equation}
with $m_{\chi}$ the WIMP mass, $m_r$ the WIMP-nucleus reduced mass, $\rho_0$   the local Dark Matter density, $\sigma_0$   the
WIMP-nucleus elastic scattering cross section, $F(E_r)$  the form factor  (using the axial expression from \cite{lewin}),  
$v_{\textrm{min}}$ the   minimal WIMP velocity required to produce a
nuclear recoil of energy $E_r$ and $\hat{q}$ the direction of the recoil momentum. 
Finally, $\hat{f}(v_{\textrm{min}},\hat{q})$ is the three-dimensional Radon transform of the WIMP 
velocity distribution $f(\vec{v})$. As the Radon transform is a linear application, one can simply add the host halo and the Dark Disk contribution to the directional event
rate. The angular distribution $dR/d\Omega_r$ is thus obtained by integrating the double-differential spectrum over the 
energy range chosen to be $E_r = [5, 50]$ keV.\\

There are several approaches to exploit the forthcoming directional data.  Either the data analysis may aim at 
rejecting the isotropy  hypothesis \cite{morgan1,morgan2,copi1,copi2,copi3,green1,green2} or at discovering galactic 
Dark Matter with a high significance \cite{billard.profile} via a profile likelihood ratio test statistic. 
In the following,  we investigate the effect of a Dark Disk contribution on the expected significance of a 
discovery of Dark Matter with directional detection. To do so,   
we use a twofold approach, following \cite{billard.profile,morgan2}.\\
First, we use a profile likelihood ratio test statistic, as presented in \cite{billard.profile} and briefly recalled hereafter for the reader's
convenience. In order to remain  independent from the background energy spectrum modelling, only the directional information is considered in the
following, {\it i.e.} the angular distribution of the recoiling events $dR/d\Omega_r$. 
Noting $\sigma_p$ the WIMP-proton cross section and $R_b$ the background rate, the likelihood function is given by,
\begin{equation}
\mathscr{L}(\sigma_p,R_b)  = \frac{(\mu_s + \mu_b)^N}{N!}e^{-(\mu_s + \mu_b)}\ \nonumber \\
  \times\prod_{n = 1}^{N} \left[ \frac{\mu_s }{\mu_s + \mu_b} S(\vec{R}_n)  + \frac{\mu_b }{\mu_s + \mu_b}B(\vec{R}_n)\right ]\ 
 \label{eq:likelihood}
 \end{equation}
where  $\mu_b = R_b\times\xi$ and $\mu_s$ 
corresponds to the number of expected background and WIMP events respectively, where $\xi$ corresponds to the exposure.
 $N$ is the total number of observed events, $\vec{R}_n$
 refers to the    
direction  of each event while the 
functions $S$ and $B$ are the directional event rate $dR/d\Omega_r$ of the WIMP  and background 
  events respectively. Following recent studies on the angular distribution of muon-induced neutrons \cite{Mei:2005gm}, the background angular distribution $B$ is assumed 
  to be isotropic in the galactic rest frame.
  Note that, contrary to a previous work \cite{billard.profile} the astrophysical uncertainties are not 
  taken into account in the estimation of the significance to allow fair comparison between the two statistical approaches.
   Only the background rate is taken as a nuisance parameter.\\
In a frequentist approach, the significance of a new process is commonly estimated by using the profile likelihood ratio 
test \cite{Cowan:2010js}. It corresponds to a
hypothesis test of the null hypothesis $H_0$ (background only) against the alternative $H_1$ which includes both background and signal.
As discussed in \cite{Cowan:2010js} the test statistic in the case of a discovery is defined as follows:
\begin{equation}
\rm q_0 = \left\{
\begin{array}{rrll}
\rm & -2\ln\lambda(0)	&	\ \hat{\sigma_p} > 0 \\
\rm & 0  		& 	\ \hat{\sigma_p} < 0
\end{array}\right.
\end{equation}
with, 
\begin{equation}
\lambda(0) = \frac{\mathscr{L}(\sigma_p = 0,\hat{\hat{R_b}})}{\mathscr{L}(\hat{\sigma_p},\hat{R_b})}
\end{equation}
Hence, a  large value of $q_0$ implies a large discrepancy between 
the two hypothesis which is in favor of a discovery ($H_1$).  As $f(q_0 \mid H_0)$ follows a $\chi^2_1$ distribution,  the discovery significance $Z$ 
is simply defined as $Z = \sqrt{q^{\rm obs}}$, in units of $\sigma$ \cite{Cowan:2010js}.\\

The second approach, first introduced by B. Morgan {\it et al.} \cite{morgan2}, is based on a generic test of isotropy following the mean recoil deviation $\langle \cos\theta
\rangle$ such as:
\begin{equation}
\langle \cos{\theta} \rangle = \frac{1}{N}\sum^N_{i=1}{\cos{\theta_i}}
\end{equation}
where $\theta_i$ is the $i^{th}$ angle between the recoil and the Cygnus direction, and $N$ is the number of
measured recoils. Note that this test, as well as the previous one, is by definition coordinate system dependent as the main recoil direction
$(\ell,b)$ (see \cite{billard.disco}) is not considered here as a fitting parameter.\\
Eventually, one can evaluate the significance of an observed anisotropy by computing the distributions of $\langle\cos\theta\rangle$ for
both $H_0$ corresponding to the background (isotropic) and $H_1$ the alternative. It is worth noticing that the use of the 
 variable $\langle\cos\theta\rangle$ is particularly interesting in the case of directional detection of Dark Matter as the expected signal should exhibit a dipole feature
 hence maximizing the deviation between $H_0$ and $H_1$.

\section{Influence of a co-rotating Dark Disk}
\label{sec:disco}

In order to investigate the effect of a Dark Disk component on the expected significance of a directional dark matter detection, 
we  allow for a wide range  on the Dark Disk parameters, see eq.~\ref{eq:ddparam}, and we evaluate, for each configuration, 
the expected significance for a 30 kg.year MIMAC-like detector. We highlight the fact that for a co-rotating Dark Disk to contribute to the data, 
the energy threshold must be low and/or the WIMP mass large.  For concreteness, we present a case study for a 50 GeV/c$^2$ WIMP 
mass and a total of 100 WIMP events. Figure~\ref{fig:dispersion} (left) presents the mean significance E(Z) as a function of $V_{DD}$, the rotation velocity of the Dark Disk at 
Solar radius. The black dashed line corresponds to the no Dark Disk case. The result is then presented for various values  
of the relative density  $\rho_{DD}/\rho_{H}$. The general feature is that the mean significance is
decreasing when increasing the co-rotating velocity of the Dark Disk at Solar radius as it results in a loss of 
directionality. This effect is even stronger when increasing the Dark Disk contribution, {\it i.e.} for large values of the relative velocity $\rho_{DD}/\rho_{H}$. 
Interestingly, a co-rotating Dark Disk can boost the mean significance of a Directional Dark Matter detection. Indeed, for a
velocity dispersion $\sigma_{DD} = 85$ km/s and a WIMP mass of 50 GeV/c$^2$, one can see that for rotation velocity $V_{DD} \leq 140$ 
km/s and for any relative density, the mean significance obtained is greater than the one obtained in the no Dark Disk case.
This enhancement of the significance at low rotation velocities can be explained by the fact that the Dark Disk is a structure colder  than the host halo, {\it i.e.} has a smaller 
velocity dispersion, implying an even more anisotropic recoil angular distribution. Hence,  a co-rotating Dark Disk will not
necessarily degrade the expected performance of directional detection. Of course, for a perfectly 
co-rotating Dark Disk ($V_{DD} = v_\odot = 220$ km/s), whatever the velocity dispersion, the recoil angular distribution induced by the 
Dark Disk is necessarily isotropic. Only the contribution of the Dark Disk to the total number of events will change.\\
Figure~\ref{fig:dispersion} (right) presents  the mean significance as a function of $V_{DD}$. For any value of the 
velocity dispersion $\sigma_{DD}$, the mean significance is continuously 
decreasing with the rotation velocity of the Dark Disk. However, the case $\sigma_{DD} = 35$ km/s (red solid line) 
tends to the no Dark Disk limit due to the
fact that the contribution to the total number of WIMP events from the Dark Disk falls quickly to zero for $V_{DD} > 140$ km/s. 
This also explains the rapid decrease of the significance enhancement in the range $0 - 100$ km/s. For larger velocity dispersions, 
the mean significance does not tend to the no Dark Disk limit as the contribution of the Dark Disk to the total number of events 
remains non negligible. Interestingly, one may note that the range of the values of
 $V_{DD}$ inducing an enhancement of the significance depends strongly on $\sigma_{DD}$. Indeed, 
 for large values of the 
 velocity dispersion, the Dark Matter signal gets closer to an isotropic distribution. 
 This observation implies that lower is the velocity
 dispersion, larger is the range in values of $V_{DD}$ allowing for a boost of the directional signature. 
 As a conclusion, larger is the velocity dispersion of the Dark Disk, weaker is the directional discovery significance, 
 except for the case of $\sigma_{DD} = 35$ km/s as discussed above. However, note that the value of $\sigma_{DD} = 141$ km/s
 is extremely large with respect to the recent results from N-Body simulations. Hence, for a 50 GeV/c$^2$ WIMP mass, one could expect that a co-rotating Dark Disk could 
 have a positive, though small ($\sim$ 10\%), effect on the directional detection of Dark Matter.

\begin{figure*}[t]
\begin{center}
\includegraphics[scale=0.5,angle=0]{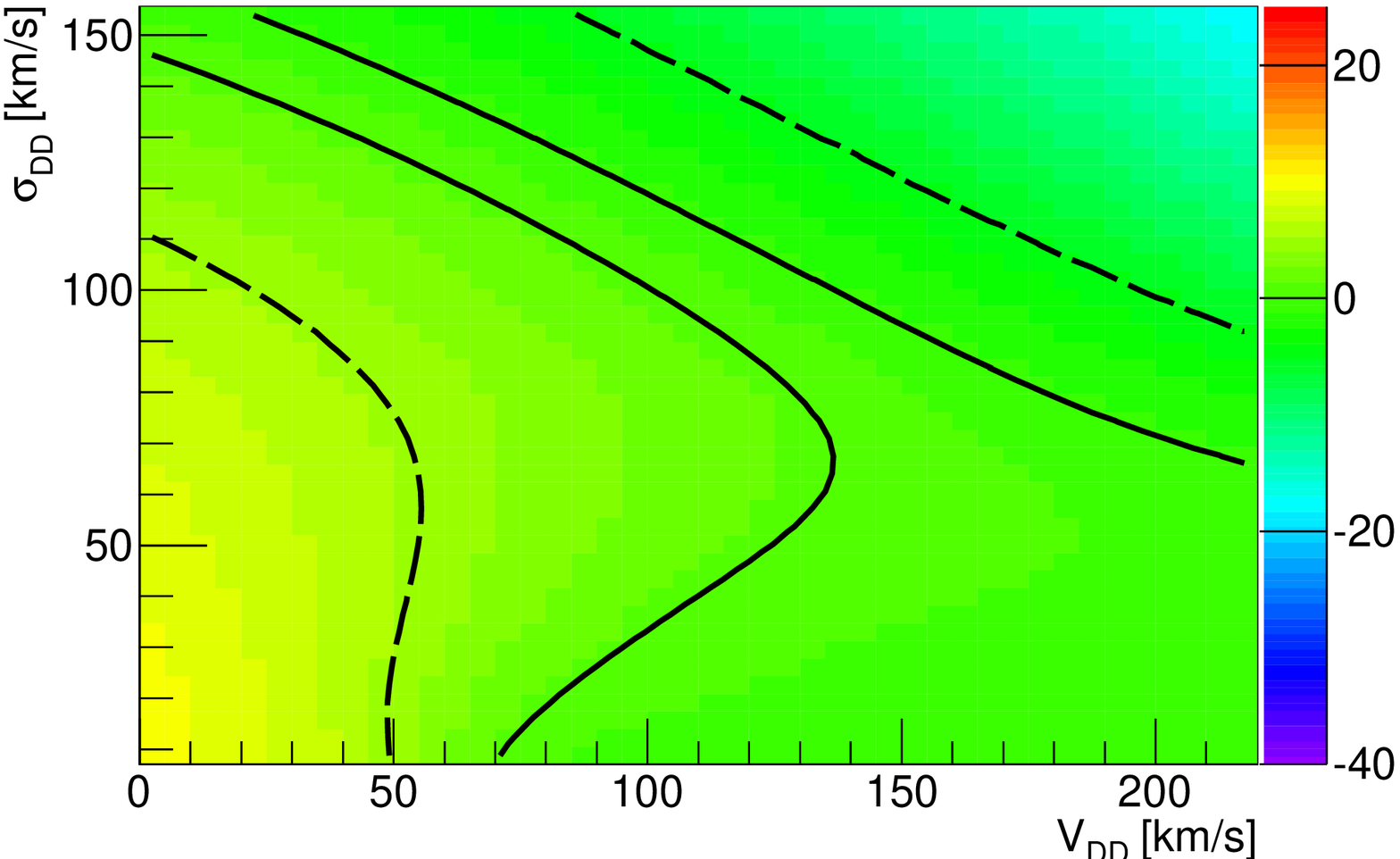}
\includegraphics[scale=0.5,angle=0]{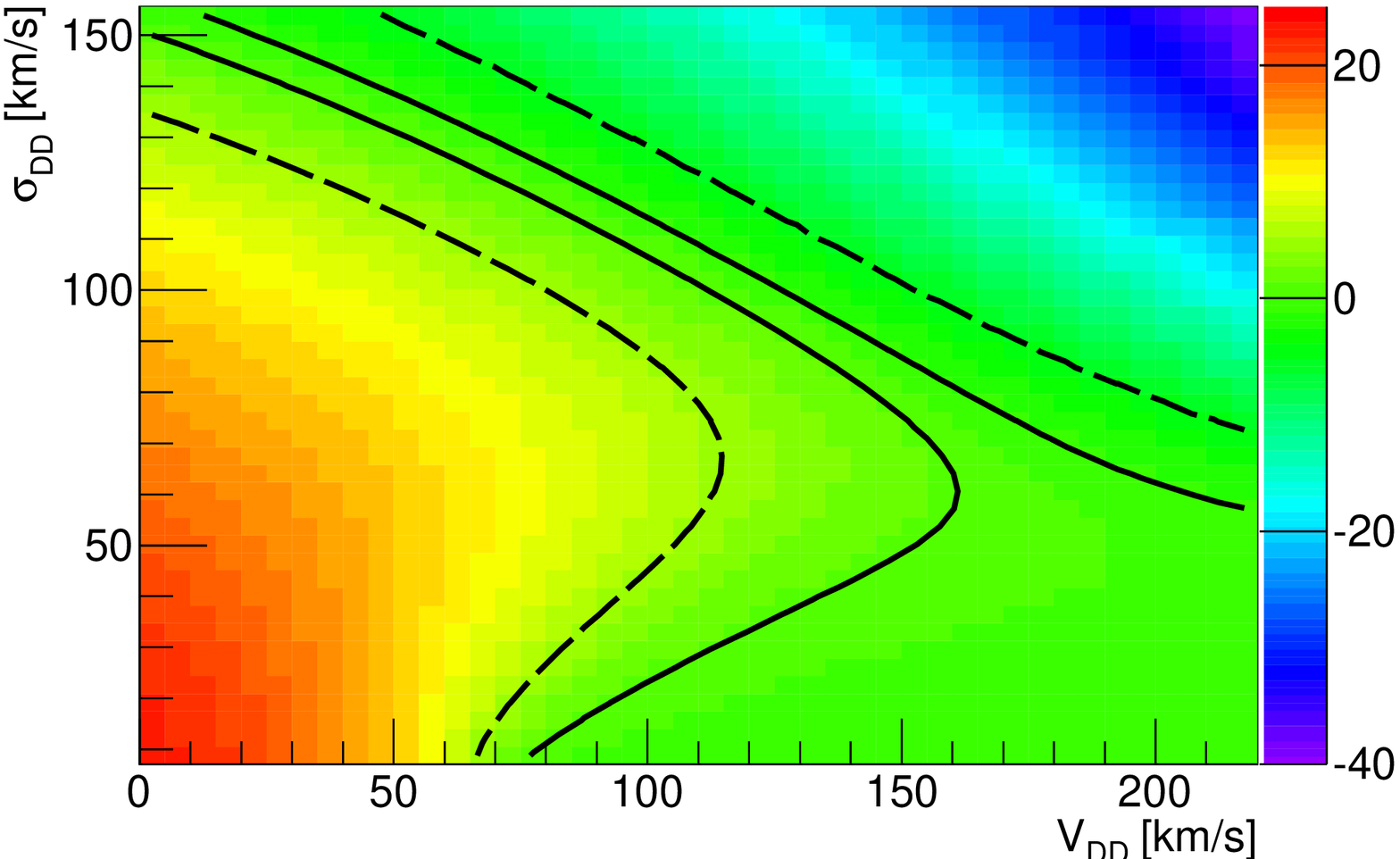}
\caption{ Relative asymmetry $\mathscr{A}$ in the plane ($V_{DD}, \sigma_{DD}$) for a relative density $\rho_{DD}/\rho_H = 1/3$ (top) 
and $\rho_{DD}/\rho_H = 1$ (bottom).
The solid and dashed lines correspond to isocontours of a relative asymmetry equal to $\pm 1$\% and $\pm 5$\% respectively.  
These studies have been done for 50 GeV/c$^2$ WIMP mass.} 
\label{fig:dispersion2}
\end{center}
\end{figure*}

For completeness, we studied the evolution of the modifications of the angular distribution $dR/d\Omega_r$ for various Dark Disk parameter values. For this purpose, we
defined the relative asymmetry $\mathscr{A}$ as:
\begin{equation}
\mathscr{A} = \frac{\langle\cos\theta\rangle - \langle\cos\theta\rangle_{H}}{\langle\cos\theta\rangle_{H}}
\label{eq:a}
\end{equation}
where $\langle\cos\theta\rangle_{H}$ corresponds to the mean recoil deviation obtained in the no Dark Disk case ($\rho_{DD} = 0$). Note that considering the standard halo
model and a WIMP of 50 GeV/c$^2$, we found $\langle\cos\theta\rangle_{H} \simeq 0.51$. Figure~\ref{fig:dispersion2} presents the relative asymmetry $\mathscr{A}$ 
in the plane ($V_{DD}, \sigma_v^{DD}$) for a relative density $\rho_{DD}/\rho_H = 1/3$ (left) and $\rho_{DD}/\rho_H = 1$ (right). One may notice that there are three
different regions : no effect (the 1\% region), a directional discovery  enhancement region (low $V_{DD}$) and a region for which the Dark Disk weakens the directional
signature (high $V_{DD}$ together with a high $\sigma_{DD}$ value). The relative density only affects the amplitude of $\mathscr{A}$, note that the latter spans the range
[-18,10] for $\rho_{DD}/\rho_H = 1/3$ and [-40,23]  for $\rho_{DD}/\rho_H = 1$. This also affects the area of the no effect region which 
 decreases with increasing value of $\rho_{DD}/\rho_{H}$. Interestingly, one can notice that most of the Dark Disk models suggested by N-Body simulations lie in the no effect
 region. Only extreme, yet unrealistic, Dark Disk models may affect significantly the directional signature. It corresponds to the case when both the co-rotational velocity and the velocity dispersion are
 high.\\

This result is in good agreement with previous work \cite{Green:2010gw} on the effect of a Dark Disk component on directional detection reach  which led to the following conclusion. 
There is only a  small variation, with respect to the no Dark Disk case, in the number of WIMP events required to reject isotropy (at 95\% confidence in 95\% of experiments) 
or to reject the median direction being random (at 95\% confidence in 95\% of experiments). 
Note that the study has been done for a Sulfure detector (a DRIFT-like one) assuming a 20 keV energy threshold and  no background.\\

\begin{figure*}[t]
\begin{center}
\includegraphics[scale=0.5,angle=0]{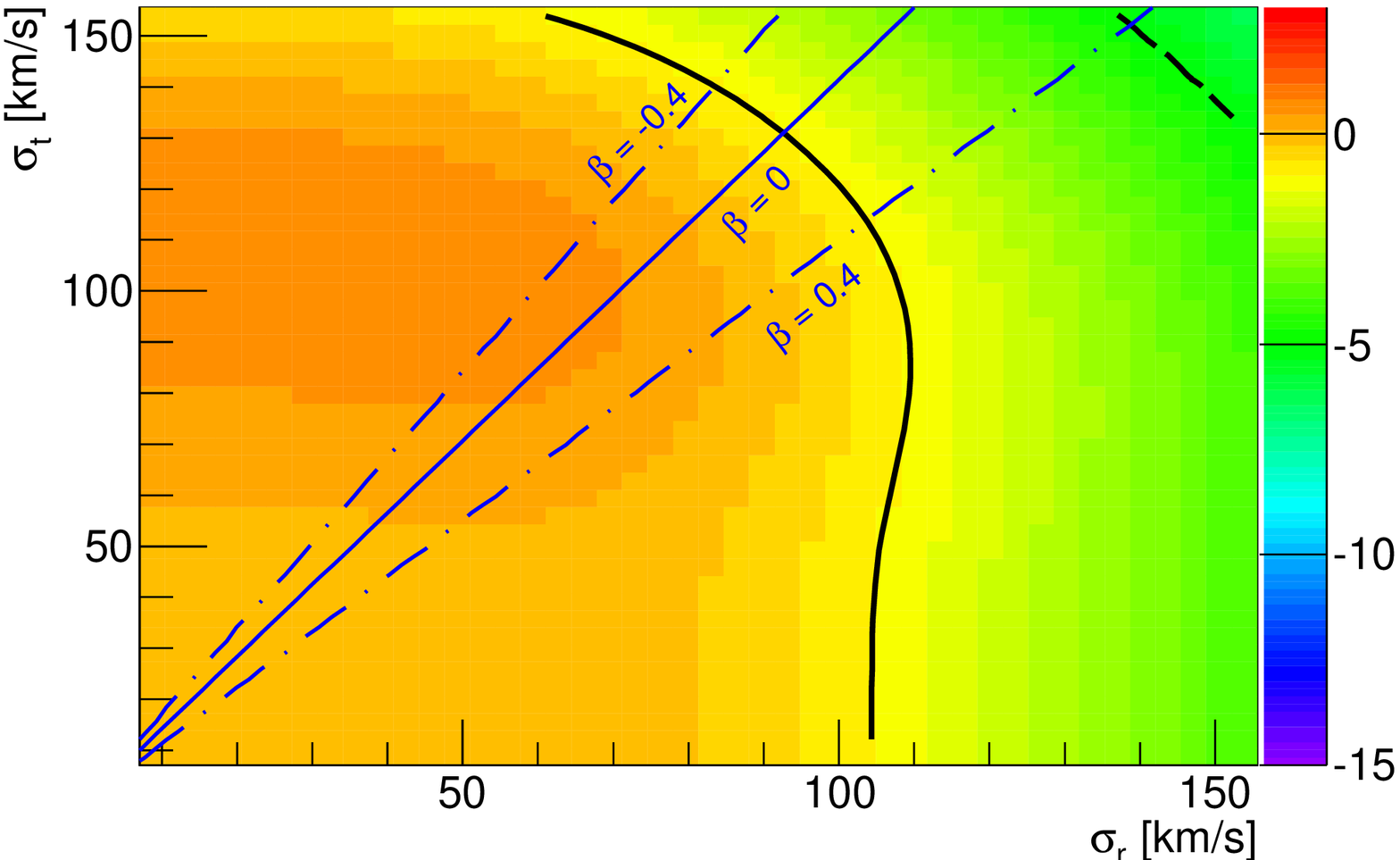}
\includegraphics[scale=0.5,angle=0]{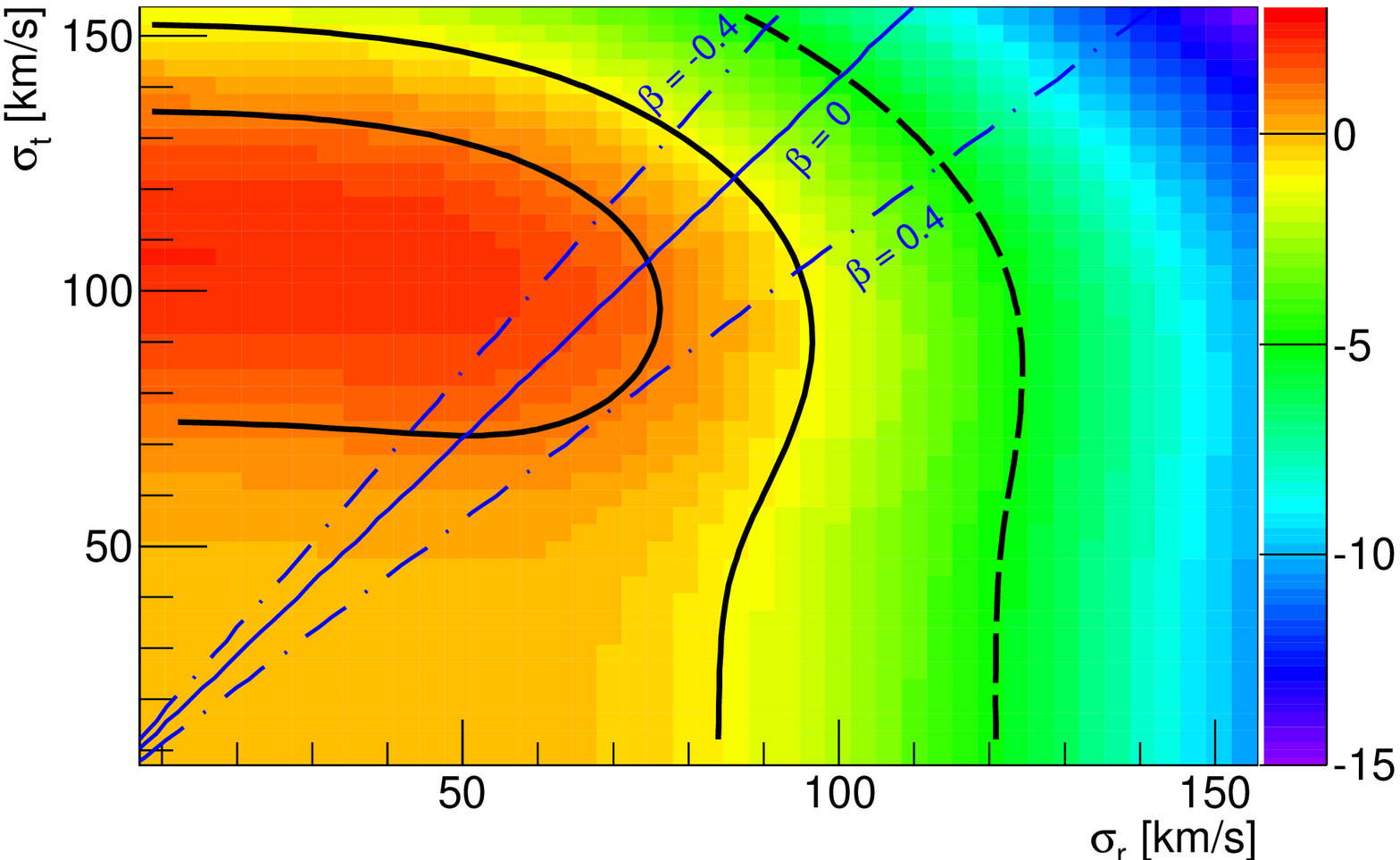}
\caption{Relative asymmetry $\mathscr{A}$ in the plane ($\sigma_{r}, \sigma_{t}$) for a relative density $\rho_{DD}/\rho_H = 1/3$ (top) 
and $\rho_{DD}/\rho_H = 1$ (bottom). The solid and dashed lines correspond to isocontours of a relative asymmetry equal to $\pm 1$\% and $\pm 5$\% respectively.  
These studies have been done for a  WIMP mass  of 50 GeV/c$^2$ and a co-rotational velocity $V_{DD} = 150 \ km/s$.} 
\label{fig:dispersion3}
\end{center}
\end{figure*}

So far, we have focused on an isotropic Maxwellian distribution for the Dark Disk particles.  
However, as shown in \cite{Ling:2009cn}, the dark disk itself may exhibit
anisotropic features in its velocity distribution. One way to investigate its
velocity dispersion tensor using current experimental data, is to look at the stellar thick disk. However, comparison with  the observed values of the velocity dispersions of the stellar thick disk may be misleading as the Dark Disk 
anisotropy depends strongly on the merger properties such as infall inclinations. Nevertheless, evidence in favor of a departure from 
isotropy comes from full cosmological hydrodynamics simulations \cite{Ling:2009cn}.\\ 
To study the effect of an anisotropic Dark Disk,  
we evaluate the value of the relative asymmetry $\mathscr{A}$ (eq.~\ref{eq:a}) as a function of the radial dispersion $\sigma_r$ and 
the tangential one, defined as $\sigma_t^2=\sigma_y^2+\sigma_z^2$. Figure \ref{fig:dispersion3} presents the 
relative asymmetry $\mathscr{A}$ in the plane ($\sigma_{r}, \sigma_{t}$) for a relative density $\rho_{DD}/\rho_H = 1/3$ (top) 
and $\rho_{DD}/\rho_H = 1$ (bottom). These studies have been done for a  
WIMP mass  of 50 GeV/c$^2$ and a co-rotational velocity $V_{DD} = 150 \ km/s$. 
For convenience the isovalues of the anisotropy parameter $\beta = 1 - \sigma_t^2/2\sigma_r^2$ are indicated. 
Note that a positive and a negative value of $\beta$ refer to a radially and tangentially 
anisotropic velocity distribution respectively.
First, it can be noticed that, for a fixed value 
of $\sigma_t$, the relative asymmetry decreases with increasing $\sigma_r$, 
{\it i.e.} perpendicularly to the detector motion direction, as the WIMP flux is becoming more isotropic 
in the detector frame, without enhancing the Dark Disk contribution to the data (see discussion above). 
For a fixed value of $\sigma_r$, due to the Earth rotation along the $(Oy)$ axis, a larger dispersion along this axis 
will mostly boost the Dark Disk contribution to the number of WIMPs events while keeping a strong anisotropy, if $\sigma_t$ is 
not too large. Hence, there is an optimal point above which, increasing $\sigma_t$ and hence both $\sigma_y$ and $\sigma_z$ starts to make the flux sufficiently anisotropic to weaken the directional signal.
Eventually it should be highlighted that for any departure from isotropy, the effect on the relative asymmetry remains small, -15\% at the very
most and in the extreme cas of a relative density of $\rho_{DD}/\rho_H = 1$. On its own, the effect of the Dark Disk anisotropy   is small compared to the influence coming from the standard parameters such as the one previously studied ($\sigma_{DD}$ and $V_{DD}$).\\

\begin{figure}[t]
\begin{center}
\includegraphics[scale=0.5,angle=0]{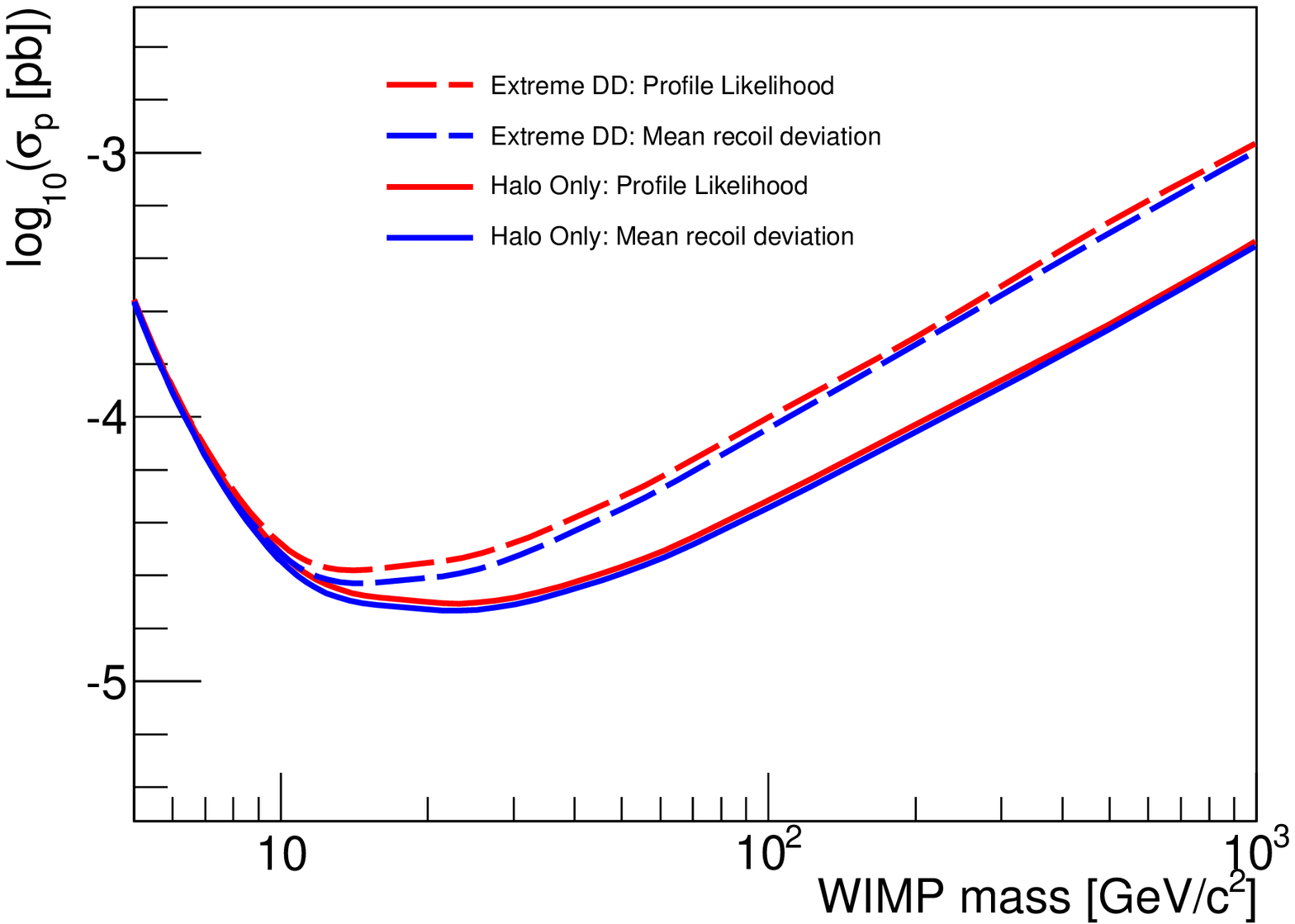}
\caption{Discovery limit in the ($m_\chi,\log_{10}(\sigma_p)$) plane corresponding to two Dark Matter models: standard halo model only (solid lines) and 
 and with an extreme Dark Disk model contribution  \{$\rho_{DD}/\rho_h = 1$, $V_{DD} = 220$ km/s, $\sigma_{DD} = 106$ km/s\} (dashed lines). 
 We compute the directional reach, {\it i.e} the  
lower bound of the 3$\sigma$ discovery region at 90\% CL, for the two approaches: profile likelihood (red lines) and mean recoil deviation (blue lines).} 
\label{fig:ProspectsFinal}
\end{center}
\end{figure}

We evaluate the effect of the Dark Disk contribution to the Dark Matter 
reach of upcoming directional detectors. Following \cite{billard.profile}, we compute the directional reach in the $(m_{\chi}, \sigma_p)$ plane, {\it i.e} the  
lower bound of the 3$\sigma$ discovery region at 90\% CL for the two approaches: profile likelihood (red lines) and mean recoil deviation (blue lines).
 Figure \ref{fig:ProspectsFinal} presents  the  
discovery limit in the ($m_\chi,\log_{10}(\sigma_p)$) plane corresponding to two Dark Matter models: standard halo model only (solid lines) and 
 and with an extreme Dark Disk model contribution  \{$\rho_{DD}/\rho_h = 1$, $V_{DD} = 220$ km/s, $\sigma_{DD} = 106$ km/s\} (dashed lines). \\
 The conclusion of this study is twofold. First, we found that for both statistical tests, the effect of an extreme Dark Disk is only mild. Indeed, the directional reach is
 only degraded by a factor of 3 at high WIMP masses and not affected for light WIMP. Second, we found that the two statistical tests give similar results with a maximal
 deviation of a few percent. However, it is worth emphasizing that their intepretation differ as the profile likelihood method favors
  the background plus signal hypothesis ($H_1$)
 whereas the mean recoil deviation method rejects the isotropy hypothesis.

\section{Conclusion}
A co-rotating Dark Disk, as predicted by recent N-Body simulations,  might contribute   
(10\%-50\%) to the local Dark Matter density, with a potentially  dramatic effect on directional detection. 
In this letter, we have evaluated the effect of  Dark Disk model on the discovery potential 
of upcoming  directional detectors.  We conclude that, if a co-rotating Dark Disk is present in our Galaxy and has the  properties  predicted by N-Body simulations \cite{nezri}, 
 the discovery potential of directional detection would be strictly unchanged.  Only an extreme and unrealistic Dark Disk model (high co-rotational velocity and high velocity dispersion) 
 might affect significantly the Dark Matter reach of upcoming directional detectors, by increasing the discovery limit by a 
 factor of three  at high WIMP mass ($m_{\chi} \sim 1000$ GeV/c$^2$). Additionally, we also have shown that anisotropic features in the Dark Matter velocity distribution of the Dark Disk will only have a small effect on the expected directional signal. Hence, according to our results we believe that the possibility of the existence of a co-rotational Dark Disk in our galaxy shouldn't be a threat for upcoming directional detection experiments.\\
 Interestingly, note that even if the impact of Dark Disk contribution to the local Dark Matter distribution only mildly affects the discovery 
 potential of directional detection, it may
 significantly affect the mass and cross section determination \cite{billard.ident}. Indeed, as explained in \cite{Green:2010gw}, 
 WIMP events arising from the Dark Disk contribution will induce an
 excess at low recoil energies which can lower the estimation of the WIMP mass when considering a standard halo model. 
 As outlined in  \cite{Lee:2012pf}, the presence of a Dark Disk restricts the ability to constrain 
 the Dark Matter parameters (both from the halo and particle physics). Of course, a measurement of the parameters of the Dark Disk itself remains challenging with the exposure 
 of the next generation of directional detectors (30 kg.year). This highlights  the fact that even if a co-rotating Dark Disk is not a threat to the discovery potential of directional detection, 
 it has to be characterized in order to consistently constrain the Dark Matter properties.






\bibliographystyle{elsarticle-num}
\bibliography{<your-bib-database>}







\end{document}